\documentclass[12pt,a4paper,notitlepage]{article}
\usepackage[utf8]{inputenc}
\usepackage[english]{babel}
\usepackage[T1]{fontenc}
\usepackage{hyperref}
\usepackage{amsmath}
\usepackage{amsfonts}
\usepackage{stmaryrd}
\usepackage{amssymb}
\usepackage{color} 
\usepackage{caption}
\usepackage[toc,page]{appendix}
\usepackage{array,multirow,makecell}
\setcellgapes{1pt}
\makegapedcells
\newcolumntype{R}[1]{>{\raggedleft\arraybackslash }b{#1}}
\newcolumntype{L}[1]{>{\raggedright\arraybackslash }b{#1}}
\newcolumntype{C}[1]{>{\centering\arraybackslash }b{#1}}

\newtheorem{proposition}{Proposition}
\newtheorem{proof}{Proof}

\usepackage{enumerate}
\usepackage{subfig}

\newcommand{\vp}{{\varphi}}

\newcommand{\beq}{\begin{equation}}
\newcommand{\eeq}{\end{equation}}
\newcommand{\bea}{\begin{eqnarray}}
\newcommand{\eea}{\end{eqnarray}}
\definecolor{mygray}{gray}{0.3}
\newcommand{\eproof}{\end{proof}}
\newcommand{\bproof}{\begin{proof}}

\newcommand{\bes}{\begin{eqnarray}}
\newcommand{\ees}{\end{eqnarray}}

\newcommand\restr[2]{{
  \left.\kern-\nulldelimiterspace 
  #1 
  \vphantom{\big|} 
  \right|_{#2} 
  }}

\usepackage{multicol}
\usepackage{amssymb}
\usepackage{graphicx}

\makeatletter

\begin{document}


\vspace{20pt}

\begin{center}

{\LARGE\bf  Energy momentum tensor for translation invariant renormalizable noncommutative field theory\\ 
\vspace{5pt}  }
\vspace{15pt}

{\large Ezinvi Balo\"{\i}tcha $^{a}$\footnote{ezinvi.baloitcha@cipma.uac.bj},  Vincent Lahoche$^{a}$\footnote{vincent.lahoche@labri.fr}  and Dine Ousmane Samary$^{a,c}$\footnote{dine.ousmane.samary@aei.mpg.de}}

\vspace{15pt}
a)\,\,{\sl Facult\'e des Sciences et Techniques, International Chair in Mathematical Physics and Applications (ICMPA-UNESCO Chair), University of Abomey-Calavi, 072B.P.50, Cotonou, Republic of Benin\\
\vspace{0.5cm}
b)\,\,{\sl LaBRI, Univ.Bordeaux
351 cours de la Lib\'eration, 33405 Talence, France, EU}\\
\vspace{0.5cm}
c)\,\, {\sl Max Planck Institute for Gravitational Physics, Albert Einstein Institute, Am M\"uhlenberg 1, 14476, Potsdam, Germany}
\vspace{0.5cm}

}
\vspace{10pt}

\begin{abstract}
In this paper, we derive the energy momentum tensor for the translation invariant  noncommutative Tanasa et al scalar field model. The Wilson regularization procedure is used to improve this tensor and  the local conservation property is recovered.
The same question is addressed in the case where the Moyal star product is deformed including the tetrad fields. It provides with an extension of the recent work [J. Phys. A: Math. Theor. 43 (2010) 155202], regarding the computation and properties of the Noether currents to the  renormalizable models.
\end{abstract}

\end{center}

\noindent  Pacs numbers:  02.40.Gh, 11.10.Nx
\\

\setcounter{footnote}{0}

\section{Introduction}
Noncommutative (NC) geometry and its applications to quantum field theory (QFT) namely NCQFT  receives an increasing attention this two decades due to the advent of the class of renormalizable actions \cite{Douglas:2001ba}-\cite{Tanasa:2008bt}. The NCQFT arises as a scenario for the Planck scale behavior of physical theories, at which the non-locality of interactions has to appear and break down the notion of continuous spacetime \cite{Doplicher:1994tu}-\cite{Majid:1999tc}.
It is most often performed over a Moyal space $\mathbb{R}_\theta^d$. This space is the deformation of $d$-dimension Euclidean space $\mathbb{R}^d$ endowed with a constant  Moyal product of functions:
\bea\label{starorigine}
(f\star g)(x)={\rm \bf m}\Big\{e^{i\frac{\theta^{\rho\sigma}}{2}\partial_\rho\otimes \partial_\sigma}f(x)\otimes g(x)\Big\},\quad x\in \mathbb{R}^d,\quad f,g\in C^{\infty}( \mathbb{R}^d).
\eea
where ${\rm\bf m}(f\otimes g)=f\cdot g$, and such that the coordinates functions $x^\sigma$ and $x^\rho$ satisfied the commutation relation
\bea
[x^\rho, x^{\sigma}]_\star=i\theta^{\rho\sigma}.
\eea
 $\theta^{\rho\sigma}$ is a skew symmetric constant tensor and elements have the dimension of length square. 
It is possible to construct the NCQFT in a nontrivial background metric, generally by imposing the non-constant deformation matrix $\theta^{\rho\sigma}=\theta^{\rho\sigma}(x)$, which naturally results in the difficulty of finding a suitable explicit closed Moyal-type formula \cite{Aschieri:2012vc}-\cite{Aschieri:2005zs}. In the context of a dynamical NC field theory this can be realized by replaced the vector field $\partial_\mu$ on the tangent space $T_x\mathbb{R}^d_\theta$ by  $X_a=e_a^\mu(x)\partial_\mu$, where the tetrad $e_a^\mu(x)$  is a tensor depending on the coordinate functions. The generalized Moyal star-product becomes
\bea\label{prot}
(f\star g)(x)={\rm \bf m}\Big\{e^{i\frac{\theta^{ab}}{2}X_a\otimes X_b}f(x)\otimes g(x)\Big\},\quad x\in \mathbb{R}^d,\quad f,g\in C^{\infty}( \mathbb{R}^d).
\eea
In general case the vector field $X_a$ are noncommutative, with respect to the Lie bracket ``$[\cdot,\cdot]$". The particular condition $[X_a,X_b]=0$ results in the constraints on the tetrad $e_a^\mu$ and leads to the definition of one new field $\vp^a$ such that  the inverse $e_\nu^a$ of $e_a^\nu$ is  proportional to $\partial_\nu \vp^a$. Since $X_a\vp^b=\delta_a^b$, the field $\vp^b$ can be viewed as  new coordinates along the  $X_a$ directions and therefore will be taking into account in the redefinition of the functional action \cite{Aschieri:2008zv}. The Moyal space $\mathbb{R}_\theta^d$ of this type becomes curved with the background metric ${\rm g}_{\mu\nu}=e_\mu^a e_\nu^b\delta_{ab}$.  Let us mention that the commuting vector field $X_a$ ensures the associativity of the star product \eqref{prot}.
 But the loss of the associativity property becomes evident in the general case where $[X_a,X_b]\neq 0$. Nevertheless, this   property maybe satisfied in a space with a nearly Euclidean metric in which  it is natural
to choose a tetrad field $e_a^\mu(x)$ that lies nearly along the coordinate
axes  $e_a^\mu(x)=\delta_a^\mu+\omega_a^\mu(x)$ where $\omega_a^\mu(x)$ is a coordinate dependent small quantity to be determined \cite{Hounkonnou:2010yk}-\cite{Hounkonnou:2011zz}. 

The basis problem, which has accompanied the development of NC field theory is the UV-IR mixing in the perturbation computation \cite{Micu:2000xj}.  This pathology maybe  solved by introducing in the scalar field action, i.e. the $\vp_\star^4$-model, the so called Grosse-Wulkenhaar (GW) harmonic term \cite{Gurau:2005gd}-\cite{Grosse:2004yu}. The GW model breaks the $U(N)$ symmetry invariance in the IR  regime but is asymptotically  safe in the UV regime. The model is also non-invariant under the translation and rotation of spacetime. The only know invariance satisfied by the model is the so called Langman-Szabo duality \cite{Langmann:2002cc}. The study of the symmetry consequence such as the Noether current are addressed  for the GW model by imposing a constrainte on the Euler-Lagrange (EL) equations of motion \cite{Geloun:2007zz}-\cite{Hounkonnou:2012td}. In \cite{Hounkonnou:2009qt} the same question is addressed in the case of twisted star product definition in the field theory.

Using the same idea of the  perturbative computation of the renormalization procedure, other theoretical model have been proved renormalizable. The theoretical ingredient to perform this issue is the so call multiscale analysis, developed by Rivasseau \cite{Rivasseau:1991ub}.  One of these models which we will focus in this work is the translation invariant renormalizable scalar model discovered  by
  Gurau-Magnen-Rivasseau-Tanasa (GMRT) \cite{Tanasa:2010fk}-\cite{Tanasa:2008bt}. The GMRT model comes from the NC $\vp^4$ model by  adding  a new contribution $\frac{a}{\theta^2 p^2}$ on the  propagator in the momentum space, and on which  the problem  of UV/IR mixing is  solved.  At any order in perturbation theory, the $\beta$ functions of the
model  are given \cite{Geloun:2008hr}. Despite all these interesting results,
the corresponding current derived from  the symmetry properties of the Tanasa model is not yet be given in the litteratures. Our purpose in this paper is to investigate the computation of the energy momentum tensor (EMT) of the GMRT model and study its regularization in both ordinary Moyal space and twisted Moyal space. The Wilson regularization procedure is used to recover the local conservation as given in  \cite{Gerhold:2000ik}-\cite{Balasin:2015hna}.

The paper is organized as follows. In section \eqref{sec2} we compute the EMT for translation invariant GMRT model in the ordinary Moyal space. The regularization of this tensor is also given. In the section \eqref{sec3} the same computation is performed in the case of the twisted Moyal plane. Our conclusion and remarks are given in section \eqref{sec4}.

\section{EMT  for renormalizable GMRT model in Moyal space }\label{sec2}
In this section  we derive the EL equation of motion and the EMT for the translation invariant nonlocal  functional action. The renormalization procedure described in \cite{Gurau:2008vd}  is also point out. Let us consider   the scalar field theoretic model  in which  we begin with the Lagrange density, which is a function of the field $\vp$, its first partial spacetime derivatives $\partial_\mu\vp$, and the inverse derivatives $\partial_\mu^{-1}\vp$ :
\bea\label{ac12}
S_\star[\vp]=\int\,d^dx\, \mathcal L_\star(\vp,\partial_\mu \vp,\partial_\mu^{-1}\vp),\quad \partial_\mu^{-1}\vp(x):=\int_{-\infty}^{x}\,dx'^\mu\, \vp(x'),
\eea
where $\mathcal{L}_\star$ mean that the ordinary product of function in the action $S$ is replaced by the Moyal product i.e.:
\bea\label{sanita}
\mathcal L_\star(\vp,\partial_\mu \vp,\partial_\mu^{-1}\vp)=\mathcal L(\vp,\partial_\mu\vp,\partial_\mu\partial_\nu\vp,\partial_\mu\partial_\nu\partial_\sigma\vp,\cdots \infty,\partial_\mu^{-1}\vp).
\eea  
 Before starting our investigation on the computation of the EMT,  let us provide the following important remark on the Lagrangian density \eqref{sanita}. First of all, the noncommutative fields theories are nonlocal in time and space due to an infinite number of temporal and spatial
derivatives in the  Lagrangian. This infinite  number of derivatives comes from the definition of the star product. Despite the fact that the inverse derivative $\partial_\mu^{-1}\vp$ is considered as a nonlocal contribution in the Lagrangian \eqref{sanita},  this Lagrangian remains nonlocal without this contribution.   The analysis performed in this work concerning the computation of the EMT becomes similar to what follows in the previous literatures, concerning the Noether current applied to noncommutative fields theories (see   \cite{Aschieri:2008zv},\cite{Geloun:2007zz}-\cite{Hounkonnou:2009qt},  \cite{Gerhold:2000ik}-\cite{Balasin:2015hna} and  references therein.)

Under the translation group, which transforms coordinates as: $x^\mu\rightarrow x^\mu+a^\mu$ ($a^\mu$ is a constant vector),  the field $\vp$ is then transform as
$
\vp(x)\rightarrow \vp(x)+a^\mu\partial_\mu\vp(x)
$, the variation of the action \eqref{ac12} gives
\bea
S_\star[\vp]\rightarrow S_\star[\vp]+\delta S_\star[\vp]= S_\star[\vp]+a^\mu\partial_\mu  S_\star[\vp],
\eea
where we have assumed that the field $\vp$ vanishes when $|x|$ approaches infinity.  We will show that the translation invariance of the nonlocal action $\delta S_\star[\vp]=0$, implies the existence of a conserved  current density $J_\nu$ such that:
\bea\label{eqreff}
\delta\vp\star\frac{\delta S_\star}{\delta\vp}+ \partial^\nu J_\nu=0,
\eea
 where the infinitesimal current $J_\nu$ may be expressed in terms of the EMT as 
\beq
\int d^dx \,\partial_\mu J^\mu:=a^\nu\int d^dx\partial_\mu T_{\nu}^{\,\,\mu}\,.
\eeq 
Note that the above relation  is well satisfied if $\partial^\alpha\vp\in\mathcal{S}(\mathbb{R}^d)$, $\alpha=\llbracket -2,-1 \rrbracket
\cup \mathbb{N}$, where $\mathcal{S}(\mathbb{R}^d)$ is the space of suitable Schwartzian functions.
 The variation of the action $S_\star$ is written as
\bea
\delta S_\star=\int\,d^dx\,\Big[\frac{\partial  \mathcal L_\star}{\partial\vp}\star\delta\vp+\frac{\partial \mathcal L_\star}{\partial\partial_\mu\vp}\star \partial_\mu\delta\vp+\frac{\partial \mathcal L_\star}{\partial\partial_\mu^{-1}\vp}\star \partial_\mu^{-1}\delta\vp\Big].
\eea
It turns out that  the inverse derivative which appears in \eqref{ac12} is such that:
$
\partial_\mu \partial_\nu^{-1}\vp(x)=\delta_{\mu\nu}\vp(x)
$
and $\partial_\mu(ab)=(\partial_\mu a) b+a(\partial_\mu b)$.
Then, the following identity is well satisfied:
\bea\label{eqer}
\sum_{\mu=1}^d \frac{\partial \mathcal L_\star}{\partial\partial_\mu^{-1}\vp}\star \partial_\mu^{-1}\delta\vp=
\sum_{\mu=1}^d\partial_\mu\Big[\partial_\mu^{-1}\Big(\frac{\partial \mathcal L_\star}{\partial\partial_\mu^{-1}\vp}\Big)\star \partial_\mu^{-1}\delta\vp\Big]-
 \sum_{\mu=1}^d\partial_\mu^{-1}\Big(\frac{\partial \mathcal L_\star}{\partial\partial_\mu^{-1}\vp}\Big)\star\delta\vp.
\eea
 Note that this relation comes from the Leibniz rule. 
In the rest of this work, without all confusions the repetitive {\it ``double indices''} is summed as Einstein summation. In the case of repetitive {\it ``triple indices''} or more, the Einstein summation is not satisfy. If this is not the case  we  will specify  if   this summation holds.
  Thus, the EL
equations of motion  for the Lagrangian density $\mathcal L_\star$
becomes
\bea\label{ELEM}
E_{\vp}=\frac{\partial \mathcal L_\star}{\partial\vp}-\partial_\mu\Big(\frac{\partial\mathcal L_\star}{\partial\partial_\mu\vp}\Big)-\partial_\mu^{-1}\Big(\frac{\partial\mathcal L_\star}{\partial\partial_\mu^{-1}\vp}\Big)=0,
\eea
and the conserved EMT can be derived by replacing in the relation \eqref{eqreff}$: \delta\vp$ by $-a^\mu\partial_\mu \vp$, such that 
\bea\label{Vincentac}
\int d^dx\, \Big(-a^\rho\,\partial^{\mu}T_{\mu\rho}+E_\vp\Big)=0,
\eea
where
\bea\label{tensor1}
T_{\mu\rho}=\frac{1}{2}\Big\{\frac{\partial\mathcal L_\star}{\partial\partial_\mu\vp},\partial_\rho \vp\Big\}_\star+\frac{1}{2}\Big\{\partial_\mu^{-1}\Big(\frac{\partial\mathcal L_\star}{\partial\partial^{-1}_\mu\vp}\Big),\partial_{\mu}^{-1}\partial_\rho\vp\Big\}_\star-g_{\mu\rho}\mathcal L_\star.
\eea
Consider as an example  the translation invariant noncommutative field theory \cite{Tanasa:2009pp}-\cite{Gurau:2008vd}, which is proposed to take into account the quantum corrections and to avoid the UV/IR mixing of the noncommutative scalar field theory:
\bea\label{action}
S_\star[\vp]=\int\, d^dx\, \Big[\frac{1}{2}\partial_\mu \vp\star \partial^\mu \vp+\frac{m^2}{2}\vp\star\vp
+\frac{a}{2\theta^2}\partial^{-1}_\mu\vp\star\partial^{-1}_\mu\vp+\frac{\lambda}{4!}\vp_\star^4 \Big],
\eea
 where
\bea\label{sam23}
\partial_\mu^{-1}\vp(x):=\int_{-\infty}^{x}\, d^\mu x' \vp (x'),\quad \int_{-\infty}^x \frac{\delta\vp(x')}{\delta\vp(y)} dx'=\Theta(x-y).
\eea
 $\Theta(x)$ is the Heaviside function.
The following statement holds
\begin{proposition}
The GMRT functional action \eqref{action} is invariant under spacetime translation. This symmetry implies the global conservation of the tensor \eqref{tensor1} due to the relation \eqref{Vincentac}. Moreover the tensor \eqref{tensor1}     leads to the construction of a  symmetric and 
 locally conserved EMT given by:
\bea\label{emtimprove}
T^{s,r}_{\rho\mu}&=&\frac{1}{2}\Big\{\partial_\mu\vp,\partial_\rho\vp\Big\}_\star+\frac{a}{4\theta^2}\Big\{\partial_\mu^{-1}\partial_{\mu}^{-1}\vp,\partial_\mu^{-1}\partial_\rho\vp\Big\}_\star+\frac{a}{4\theta^2}\Big\{\partial_\rho^{-1}\partial_{\rho}^{-1}\vp,\partial_\rho^{-1}\partial_\mu\vp\Big\}_\star\cr
&&-i\frac{\lambda}{4!}\theta^{\alpha\beta}g_{\alpha\rho}\Big( [\partial_\mu \vp,\vp]_\star \star' \partial_\beta( \vp\star\vp)\Big)-i\frac{\lambda}{4!}\theta^{\alpha\beta}g_{\alpha\mu}\Big( [\partial_\rho \vp,\vp]_\star \star' \partial_\beta( \vp\star\vp)\Big)\cr
&&-g_{\mu\rho}\mathcal L_\star.
\eea 
where we have introduced the new star product $\star'$, which is non-associative and commutative and is given by
\bea\label{starori}
f\star'g={\bf m}\Bigg\{\frac{\sin\Big(\frac{1}{2}\theta^{\mu\nu}\partial_\mu\otimes\partial_\nu\Big)}{\frac{1}{2}\theta^{\mu\nu}\partial_\mu\otimes\partial_\nu} (f\otimes g)\Bigg\}.
\eea 
\end{proposition}
\textit{Proof:}
The variation principle  gives the EL equations of motion (see equation  \eqref{ELEM} for more detail):
\bea\label{eqmotion1}
\frac{\delta S_\star}{\delta\vp}=0 \Leftrightarrow -\partial_\mu\partial_\mu \vp+m^2\vp+\frac{\lambda}{3!}\vp_\star^3-\frac{a}{\theta^2}\partial^{-1}_\mu\partial^{-1}_{\mu}\vp=0.
\eea
Furthermore  the EMT becomes, 
\bea\label{emtpre}
T_{\mu\rho}=\frac{1}{2}\Big\{\partial_\mu\vp,\partial_\rho\vp\Big\}_\star+\frac{a}{2\theta^2}\Big\{\partial_\mu^{-1}\partial_{\mu}^{-1}\vp,\partial_\mu^{-1}\partial_\rho\vp\Big\}_\star-g_{\mu\rho}\mathcal L_\star.
\eea
The tensor $T_{\mu\rho}$ is nonsymmetric and nonlocally conserved. Let $T^s_{\mu\rho}$  be the symmetric tensor associated to $T_{\mu\rho}$ i.e. $T^s_{\mu\rho}=(T_{\mu\rho}+T_{\rho\mu})/2$, we get
\bea\label{emttt}
T^s_{\mu\rho}=\frac{1}{2}\Big\{\partial_\mu\vp,\partial_\rho\vp\Big\}_\star+\frac{a}{4\theta^2}\Big\{\partial_\mu^{-1}\partial_{\mu}^{-1}\vp,\partial_\mu^{-1}\partial_\rho\vp\Big\}_\star+\frac{a}{4\theta^2}\Big\{\partial_\rho^{-1}\partial_{\rho}^{-1}\vp,\partial_\rho^{-1}\partial_\mu\vp\Big\}_\star-g_{\mu\rho}\mathcal L_\star.
\eea
Note that the  procedure of regularization of the EMT performed by Gerhold et al \cite{Gerhold:2000ik} can be used. Consider the star product $\star'$ given in \eqref{starori},
which satisfies the following identity:
\beq
\theta^{\mu\nu}\partial_\mu f \star' \partial_\nu g=-i[f,g]_\star.
\eeq
Then after few computation we get
\bea
\partial^\rho T^s_{\rho\mu}=\frac{\lambda}{4!}\Big[[\partial_\mu \vp,\vp]_\star,\vp\star\vp\Big]_\star=i\frac{\lambda}{4!}\theta^{\alpha\beta}\partial_\alpha\Big( [\partial_\mu \vp,\vp]_\star \star' \partial_\beta( \vp\star\vp)\Big).
\eea
The locally conserved EMT then becomes
\bea\label{emtimprove1}
T^{s,r}_{\rho\mu}&=&\frac{1}{2}\Big\{\partial_\mu\vp,\partial_\rho\vp\Big\}_\star+\frac{a}{4\theta^2}\Big\{\partial_\mu^{-1}\partial_{\mu}^{-1}\vp,\partial_\mu^{-1}\partial_\rho\vp\Big\}_\star+\frac{a}{4\theta^2}\Big\{\partial_\rho^{-1}\partial_{\rho}^{-1}\vp,\partial_\rho^{-1}\partial_\mu\vp\Big\}_\star\cr
&&-i\frac{\lambda}{4!}\theta^{\alpha\beta}g_{\alpha\rho}\Big( [\partial_\mu \vp,\vp]_\star \star' \partial_\beta( \vp\star\vp)\Big)-i\frac{\lambda}{4!}\theta^{\alpha\beta}g_{\alpha\mu}\Big( [\partial_\rho \vp,\vp]_\star \star' \partial_\beta( \vp\star\vp)\Big)\cr
&&-g_{\mu\rho}\mathcal L_\star.
\eea 
\begin{flushright}
$\square$
\end{flushright}
Note that the limit $a\rightarrow 0$ gives the EMT for the scalar field theory on Moyal space derived in \cite{Gerhold:2000ik} and \cite{AbouZeid:2001up} from which the Belifante PDE can be given. Also by adding the quantity $\frac{1}{6}(g_{\mu\rho}\square-\partial_\mu\partial_\rho)(\vp\star\vp)$ in the expression \eqref{emttt} and by setting $m=0$, we obtain the traceless EMT. The conventional tensor \eqref{emtpre} does not have finite matrix elements even to lowest order in the coupling $\lambda$. However, the modified tensor $T^I_{\mu\rho}=T_{\mu\rho}-\frac{1}{6}(g_{\mu\rho}\square-\partial_\mu\partial_\rho)(\vp\star\vp)$ has finite matrix elements to all order in $\lambda$.
The improvement term does not contribute to the divergence of the energy- momentum tensor, i.e.
\bea
\partial_\mu T^I_{\mu\rho}=\partial_\mu T_{\mu\rho}.
\eea
The global conservation of the EMT \eqref{emtimprove} implies the existence of a conserved $d$-momentum $P_\rho$ such that
\bea
\partial^0 P_\rho=\partial^0\int d^dx\, T^{s,r}_{0\mu}=0.
\eea

 Due to the presence of the deformation parameter $\theta^{\mu\nu}$ (as a constant, skewsymmetric, fixed tensor), the  Lorentz symmetry is manifestly broken. In the next section we introduce a deformation of the Moyal algebra so called dynamical Moyal algebra, in which the tensor $\theta^{\mu\nu}$ depends now on the coordinates. In this situation the Lorentz symmetry maybe restored and therefore the corresponding EMT becomes Lorentz invariant tensor.   Consider the tensor  $x_{\rho\mu}(a,\theta)$ given by
\bea\label{samss23}
x_{\rho\mu}(a,\theta)=\frac{a}{4\theta^2}\Big\{\partial_\mu^{-1}\partial_{\mu}^{-1}\vp,\partial_\mu^{-1}\partial_\rho\vp\Big\}_\star+\frac{a}{4\theta^2}\Big\{\partial_\rho^{-1}\partial_{\rho}^{-1}\vp,\partial_\rho^{-1}\partial_\mu\vp\Big\}_\star.
\eea
The indices $\rho$ and $\mu$ are not summed in the right hand side of \eqref{samss23}. Also we can use the fact that $\partial_\mu^{-1}\partial_\rho\vp=\delta_{\mu\rho}\vp$ and therefore $x_{\rho\mu}(a,\theta)$ becomes
\bea\label{alphaomega}
x_{\rho\mu}(a,\theta)&=&\frac{a}{4\theta^2}\Big\{\partial_\mu^{-1}\partial_{\mu}^{-1}\vp,\vp\Big\}_\star+\frac{a}{4\theta^2}\Big\{\partial_\rho^{-1}\partial_{\rho}^{-1}\vp,\vp\Big\}_\star\cr
&=&\frac{a}{4\theta^2}\Big\{\int\,d^\mu x\Big(\int\,d^\mu x\, \vp\Big),\vp\Big\}_\star+\frac{a}{4\theta^2}\Big\{\int\,d^\rho x\Big(\int\,d^\rho x\,\vp\Big),\vp\Big\}_\star.
\eea
Furthermore the tensor $t_{\rho\mu}$ providing from the $a$-dependence on the action \eqref{action} is written as
\bea\label{deni}
t_{\rho\mu}(a,\theta)=x_{\rho\mu}(a,\theta)-\frac{a}{2\theta^2}g_{\rho\mu} \partial_{\sigma}^{-1}\vp\star \partial_\sigma^{-1}\vp
\eea
We turn to consider this quantity as the regularization contribution  of the EMT for the scalar $\vp_\star^4$ theory  due to the presence  of  a term allowing renormalization of the action \eqref{action} i.e. $\partial^{-1}_\mu\vp\star\partial_\mu^{-1}\vp$. 

\section{The EMT for the GMRT model in the generalized Moyal space}\label{sec3}
This section is devoted to the computation of the EMT of the  generalized type GMRT model.  Before defined this model we give some definitions and identities satisfied by the star product \eqref{prot}. These will be used to calculate the $\vp$ and the $\vp^a$ variation of the functional action (for more explanation see \cite{Aschieri:2008zv}).
Expanding
 the generalized $\star$-product \eqref{prot}  as follows
\begin{eqnarray}\label{starpro}
f\star g
&\equiv & e^{\Delta}(f,g)=\sum_{n=0}^\infty \frac{\Delta^n}{n!}(f,g),\quad \Delta(f,g)=\frac{i}{2}\theta^{ab}(X_{a}f)(X_{b}g),
\end{eqnarray}
allows us to defined  the four operators:
\begin{eqnarray}
T(\Delta)&=&\frac{e^{\Delta}-1}{\Delta}\qquad
S(\Delta)=\frac{\sinh(\Delta)}{\Delta}\nonumber
\\ R(\Delta)&=&\frac{\cosh(\Delta)-1}{\Delta}\mbox{ and } \widetilde{X}^{a}=\frac{i}{2}\theta^{ab}X_{b},
\end{eqnarray}
 such that  the following identities hold:
\begin{eqnarray}
f\star g &=&fg+X_{a}T(\Delta)(f,\widetilde{X}^{a}g)\\
f\star g-g\star f &=&[f,g]_{\star}=2X_{a}S(\Delta)(f,\widetilde{X}^{a}g)\\
f\star g+ g\star f&=& \{f,g\}_{\star}=2fg+2X_{a}R(\Delta)(f,\widetilde{X}^{a}g).
\end{eqnarray}
$S(\Delta)(.,\widetilde{X}.)$ is a bilinear antisymmetric operator such that
\begin{eqnarray}
 T(\Delta)(f,\widetilde{X}^{a}g)-T(\Delta)(g,\widetilde{X}^{a}f)=2S(\Delta)(f,\widetilde{X}^{a}g).
\end{eqnarray}
The integral of the form $\int\,\mbox{d}^{d}x \, (f\star g)$    is not cyclic; even with suitable boundary conditions at infinity, i.e.
\begin{eqnarray}
\int\,\mbox{d}^{d}x \, (f\star g)\neq \int\,\mbox{d}^{d}x \, (g\star f).
\end{eqnarray}
Using now the measure $ed^dx$ where $e=det(e_\mu^a)$, a cyclic integral can be defined so that, up to boundary terms:
\begin{eqnarray}\label{def36}
 \int\,e\mbox{d}^{d}x \, (f\star g)=\int\,e\mbox{d}^{d}x(fg)=\int\,e\mbox{d}^{d}x \, (g\star f).
\end{eqnarray}
From now the peculiar Euler Lagrange equations of motion can be
readily derived by direct application of the variation principle and the use of
 formulas of derivatives and variations  given  in \cite{Aschieri:2008zv} by:
\begin{eqnarray}
\delta_{\vp^{c}}e=eX_{a}(\delta\vp^{a}),\quad
 \delta_{\vp^{c}}e^{-1}=-e^{-1}X_{a}(\delta\vp^{a}),
\quad
eX_{a}(f)=\partial_{\mu}(ee_{a}^{\mu}f).
\end{eqnarray}
To compute $\delta_{\vp^{c}}$ variations, consider the functions $f$ and $g$, which  do not depend on $\vp^{c}$. It turns out that the following identity is useful:
\begin{eqnarray}\label{111}
 \delta_{\vp^{c}}(f\star g)=-(\delta\vp^{c}X_{c}f)\star
  g-f\star( \delta\vp^{c}X_{c}g)+\delta\vp^{c}X_{c}(f\star g).
\end{eqnarray}

In view of all  these considerations,  let us suppose now, and in the following, that the field theory is defined  by the  so called  the generalized  GMRT model which is  described by the functional action 
\bea\label{action1}
S_\star[\vp]&=&\int\, e d^dx\, \Big\{\frac{1}{2}\partial_\mu \vp\star \partial^\mu \vp+\frac{a}{2\theta^2}\partial^{-1}_\mu\vp\star\partial^{-1}_\mu\vp+\frac{m^2}{2}\vp\star\vp
+\frac{\lambda}{4!}\vp\star\vp\star\vp\star\vp\cr
&&+\frac{1}{2}\partial_\mu \vp_c\star \partial^\mu \vp^c+\frac{a}{2\theta^2}\partial^{-1}_\mu\vp_c\star\partial^{-1}_\mu\vp^c \Big\}\star e^{-1}\cr
&=& \int\, e d^dx\,( \mathcal{L}_\star \star  e^{-1}).
\eea
Using expression \eqref{def36}, the action \eqref{action1} can also be written as 
$
S_\star[\vp]=\int\,  d^dx\, \mathcal{L}_\star,
$
and then we can easily show that \eqref{action1} is invariant under spacetime translations. The application of the Noether method to this action which admit continuous symmetrie yields locally conserved EMT such that the following result holds
\begin{proposition}\label{propre2}
The symmetric locally conserved EMT derived from the translation invariance of the action \eqref{action1} is given by the following relation:
\begin{eqnarray}\label{energy12}
\mathcal{T}^{s}_{\nu\sigma}&=&\frac{e}{4}\Big[(\partial_{\nu}\vp)\{\partial_{\sigma}\vp,e^{-1}\}_{\star}
+(\partial_{\nu}\vp_{c})\{\partial_{\sigma}\vp^{c},e^{-1}\}_{\star}+(\partial_{\sigma}\vp)\{\partial_{\nu}\vp,e^{-1}\}_{\star}
+(\partial_{\sigma}\vp_{c})\{\partial_{\nu}\vp^{c},e^{-1}\}_{\star}\Big]\cr
&&+\frac{a}{4\theta^2}\Big[\partial_\sigma^{-1}(\partial_{\nu}\vp) \partial_\sigma^{-1}\Big(e\{\partial_\sigma^{-1}\vp,e^{-1}\}_\star\Big)+\partial_\sigma^{-1}(\partial_{\nu}\vp_c) \partial_\sigma^{-1}\Big(e\{\partial_\sigma^{-1}\vp^c,e^{-1}\}_\star\Big)\cr
&&+\partial_\nu^{-1}(\partial_{\sigma}\vp) \partial_\nu^{-1}\Big(e\{\partial_\nu^{-1}\vp,e^{-1}\}_\star\Big)+\partial_\nu^{-1}(\partial_{\sigma}\vp_c) \partial_\nu^{-1}\Big(e\{\partial_\nu^{-1}\vp^c,e^{-1}\}_\star\Big)\Big]
\cr
&&-\frac{e}{2}\Big\{ g_{\rho\sigma}e_{b}^{\rho}\Big[\mathcal{L}_{\star}\star(e^{-1}\partial_{\nu}\vp^{b})+T(\Delta)
\Big(X_{c}\mathcal{L}_{\star},\widetilde{X}^{b}(e^{-1}\partial_{\nu}\vp^{c})\Big)\Big]\cr
&&+g_{\rho\nu}e_{b}^{\rho}\Big[\mathcal{L}_{\star}\star(e^{-1}\partial_{\sigma}\vp^{b})
+T(\Delta)
\Big(X_{c}\mathcal{L}_{\star},\widetilde{X}^{b}(e^{-1}\partial_{\sigma}\vp^{c})\Big)\Big]\Big\}.
\end{eqnarray}
\end{proposition}

As mentioned in the last section the triple index appearing in  \eqref{energy12} is not  summed.
Using the fact that $\partial_\mu^{-1}\partial_\rho\vp=\delta_{\mu\rho}\vp$, the tensor $\widetilde{x}_{\nu\sigma}$ corresponds to $x_{\nu\sigma}$  given in  \eqref{alphaomega} by replacing the Moyal product \eqref{starorigine} by \eqref{prot} is
\bea
\widetilde{x}_{\nu\sigma}(a,\theta)&=&\frac{a}{4\theta^2}\Big[\vp\int\,d^\sigma x\Big(e\{\int\,d^\sigma x\,\vp,e^{-1}\}_\star\Big)+\vp_c \int\,d^\sigma\, x\Big(e\{\int\,d^\sigma x\vp^c,e^{-1}\}_\star\Big)\cr
&&+\vp\int\,d^\nu x\Big(e\{\int\,d^\nu x\vp,e^{-1}\}_\star\Big)+\vp_c\int\,d^\nu x\Big(e\{\int\,d^\nu x\vp^c,e^{-1}\}_\star\Big)\Big].
\eea
Then the tensor $\widetilde{t}_{\nu\sigma}$ providing from the $a$-dependence on the action \eqref{action1} is written as
\bea
\widetilde{t}_{\nu\sigma}(a,\theta)&=&\widetilde{x}_{\nu\sigma}(a,\theta)-\frac{e}{2}\Big\{ g_{\rho\sigma}e_{b}^{\rho}\Big[\mathcal{L}^{(a)}_{\star}\star(e^{-1}\partial_{\nu}\vp^{b})+T(\Delta)
\Big(X_{c}\mathcal{L}^{(a)}_{\star},\widetilde{X}^{b}(e^{-1}\partial_{\nu}\vp^{c})\Big)\Big]\cr
&&+g_{\rho\nu}e_{b}^{\rho}\Big[\mathcal{L}^{(a)}_{\star}\star(e^{-1}\partial_{\sigma}\vp^{b})
+T(\Delta)
\Big(X_{c}\mathcal{L}^{(a)}_{\star},\widetilde{X}^{b}(e^{-1}\partial_{\sigma}\vp^{c})\Big)\Big]\Big\},
\eea
where $$\mathcal{L}^{(a)}_{\star}=\frac{a}{2\theta^2}\Big(\partial^{-1}_\mu\vp\star\partial^{-1}_\mu\vp+\partial^{-1}_\mu\vp_c\star\partial^{-1}_\mu\vp^c\Big).$$

The rest of this paper is devoted to the prove of proposition \eqref{propre2}.
let us recall that the case where $a=0$ in \eqref{action1} is reduced to the well know scalar field theory in the litterature (see \cite{Aschieri:2008zv} for more details). Then we will focus our attention on to the $\vp$ and $\vp^c$ variation  of the quantity
\bea\label{actiontanasa}
S_{\partial}=\frac{a}{2\theta^2}\int\, e d^dx\, \Big[\partial^{-1}_\mu \vp\star \partial_\mu^{-1} \vp+\partial^{-1}_\mu\vp_c\star\partial^{-1}_\mu\vp^c \Big]\star e^{-1}.
\eea
  See Appendix for more detail.

 The $\vp$ variation of the action \eqref{action1}  gives the EL equations of motion of the field $\vp$ as
\bea\label{EL50}
E_\vp&=&-\frac{1}{2}\partial_\mu(e\{\partial^\mu \vp,e^{-1}\}_\star)-\frac{a}{2\theta^2} \partial_{\mu}^{-1}(e\{\partial^{-1}_\mu \vp,e^{-1}\}_\star)+\frac{m^2}{2}e\{\vp,e^{-1}\}_\star\cr
&&+\frac{\lambda}{4!}e\{\vp\star\vp,\{\vp,e^{-1}\}_\star\}_\star=0,
\eea
which is reduced to \eqref{eqmotion1} in the limit where $X_a\rightarrow \partial_a$. Hence, the corresponding current is
\bea\label{C51}
\mathcal K^\sigma&=&\frac{a}{2\theta^2}\Big[(\partial_\sigma^{-1}\delta\vp)\partial^{-1}_\sigma(e\{\partial_\sigma^{-1}\vp, e^{-1}\}_\star)+  e e_b^\sigma T(\Delta)(\partial_\mu^{-1}\delta\vp,\widetilde{X}^b(\{\partial_\mu^{-1}\vp, e^{-1}\}_\star))\cr
&&+2  e e_b^\sigma S(\Delta)(\partial_\mu^{-1}\vp,\widetilde{X}^b(\partial_\mu^{-1}\delta\vp\star e^{-1}))\Big]+ \frac{e\delta\vp}{2}.\{\partial^{\sigma}\vp,e^{-1}\}_{\star}\cr
&&+ee_{b}^{\sigma}\Big[T(\Delta)\Big(\delta\partial_{\mu}\vp,\frac{\widetilde{X}^{b}}{2}\{\partial^{\mu}\vp,e^{-1}\}_{\star}\Big)+S(\Delta)\Big(\partial_{\mu}\vp,\widetilde{X}^{b}(\partial^{\mu}\delta\vp\star e^{-1})\Big) \Big]\cr
&&+ee_{b}^{\sigma}\Big[\frac{m^2}{2}T(\Delta)
\Big(\delta\vp,\widetilde{X}^{b}\{\vp,e^{-1}\}_{\star}
\Big)+m^{2}S(\Delta)\Big(\vp,\widetilde{X}^{b}(\delta\vp\star
e^{-1})\Big)\Big]\cr
&&+ee_{b}^{\sigma}\Big[\frac{\lambda}{4!}T(\Delta)\Big(\delta\vp,\widetilde{X}^{b}\{\vp\star\vp,\{\vp,e^{-1}\}_{\star}\}_{\star}\Big)+\frac{\lambda}{12}S(\Delta)\Big(\vp,\widetilde{X}^{b}(\delta\vp\star\vp\star\vp\star e^{-1})\Big)
\cr
&& +\frac{\lambda}{12}S(\Delta)\Big(\vp\star\vp,\widetilde{X}^{b}(\delta\vp\star\vp\star e^{-1})\Big)+\frac{\lambda}{12}S(\Delta)\Big(\vp\star\vp\star\vp,\widetilde{X}^{b}(\delta\vp\star e^{-1})\Big)\Big],
\eea
such that
\bea\label{52}
\delta_{\vp}S_{\star}&=&\int\,d^dx\Big[\delta\vp E^\vp+\partial_\sigma\mathcal K^\sigma].
\eea
In the other hand the $\vp_c$ variation of \eqref{action1} gives the EL-equations of motion
\begin{eqnarray}\label{EL53}
 {E}^{\vp\vp^c}&=&E^{\vp^c}_{\partial}-X_{c}\vp\mathcal{E}_{\vp}+X_{c}\mathcal{L}_{\star}^{\Omega}
-\frac{1}{2}X_{c}\vp\partial_{\mu}\Big(e\{\partial^{\mu}\vp,e^{-1}\}_{\star}\Big)-e\frac{\Omega^{2}}{2}\vp X_{c}\tilde{x}.\{\tilde{x}\vp,e^{-1}\}_{\star}\cr
&&-\frac{e}{2}X_{c}\partial_{\mu}\vp.\{\partial^{\mu}\vp,e^{-1}\}_{\star}-\frac{e}{2}X_{c}\partial_{\mu}\vp_{a}.\{\partial^{\mu}\vp^{a},e^{-1}\}_{\star}-\partial_{\mu}\Big(\frac{e}{2}\{\partial^{\mu}\vp_{c},e^{-1}\}_{\star}\Big)=0
\end{eqnarray}
and the corresponding current 
\begin{eqnarray}\label{C54}
\mathcal{J}^{\sigma}&=&\mathcal{J}_\partial^{\sigma}+\mathcal{K}^{\sigma}(\delta\vp\rightarrow-\delta\vp^{c}X_{c}\vp)
+\frac{e\delta\vp^{c}}{2}X_{c}\vp.\{\partial^{\sigma}\vp,e^{-1}\}_{\star}
+\frac{e\delta\vp^{c}}{2}.\{\partial^{\sigma}\vp_{c},e^{-1}\}_{\star}\cr &&+ee_{b}^{\sigma}\Big\{-\mathcal{L}_{\star}\star(\delta\vp^{b}e^{-1})+
\delta\vp^b (\mathcal{L}_{\star} \star e^{-1})+T(\Delta)\Big(X_{c}(\mathcal{L}_{\star}),\widetilde{X}^{b}(\delta\vp^{c}e^{-1})\Big)\cr 
&&+\frac{1}{2}T(\Delta)\Big(\partial_{\mu}(\delta\vp^{c}e_{c}^{\rho})\partial_{\rho}\vp,
\widetilde{X}^{b}\{\partial^{\mu}\vp,e^{-1}\}_{\star}\Big)+S(\Delta)\Big(\partial_{\mu}\vp,
\widetilde{X}^{b}((\partial_{\mu}(\delta\vp^{c}e_{c}^{\rho})\partial_{\rho}\vp)\star
 e^{-1})\Big)\Big\}\nonumber\\
&&+\frac{1}{2}ee_{b}^{\sigma}\Big\{-T(\Delta)\Big(\delta\vp^{c}X_{c}\partial_{\mu}\vp_{a},
\widetilde{X}^{b}\{\partial^{\mu}\vp^{a},e^{-1}\}_{\star}\Big)-2S(\Delta)\Big(\partial^{\mu}\vp_{a},
\widetilde{X}^{b}((\delta\vp^{c}X_{c}\partial_{\mu}\vp^{a})\star e^{-1})\Big)\cr &&+2S(\Delta)\Big(\partial_{\mu}\vp_{a},\widetilde{X}^{b}(\partial^{\mu}\delta\vp^{a}
\star e^{-1})\Big)+T(\Delta)\Big(\partial_{\mu}\delta\vp_{a},\widetilde{X}^{b}
\{\partial^{\mu}\vp^{a},e^{-1}\}_{\star}\Big)\Big\},
\end{eqnarray}
such that 
\begin{eqnarray}\label{55}
\delta_{\vp^{c}}\mathcal{S}_{\star}&=&\int{\mbox{d}}^d x\hspace{1mm}\Big(\delta\vp^{c}\,{E}^{\vp\vp^c}+\partial_{\sigma}\mathcal{J}^{\sigma}\Big).
\end{eqnarray}

Now using the results in the previous paragraph where we studied the general
properties of the total variation   of the Lagrangian, we discuss the translation invariant symmetry of the model and compute the conserve current namely the EMT. In general, a symmetry of the action involves a certain change of variables.
Performing a  functional variation of the fields and  a coordinates transformations
\begin{eqnarray}
 \vp'(x)=\vp(x)+\delta\vp(x),\quad
\vp'^{c}(x)=\vp^{c}(x)+\delta\vp^{c}(x),\quad
x'^{\mu}=x^{\mu}+a^{\mu},
\end{eqnarray}
and by using the identity 
 $\mbox{d}^{D}x'=[1+\partial_{\mu}a^{\mu}+\mathcal{O}(a^{2})]\mbox{d}^{D}x$,
leads  to the following  variation of the action, to first order in
 $\delta\vp(x), \delta\vp^{c}(x)$ and $a^{\mu}$:
\begin{eqnarray}
\delta\mathcal{S}_{\star}&=& \int\,e\mbox{d}^{d}x\Big\{\Big{\vert}\frac{\partial x'}{\partial x}
\Big{\vert}\star(\mathcal{L'}_{\star}\star e^{-1})\Big\}-\int\,e\mbox{d}^{d}x\,
 (\mathcal{L}_{\star}\star e^{-1})\nonumber\\
&=&\int{\mbox{d}}^d x\hspace{1mm}\Big\{\delta_\vp\Big((\mathcal{L}_{\star}\star e^{-1})e\Big)
+\delta_{\vp^{c}}\Big((\mathcal{L}_{\star}\star e^{-1})e\Big)
\cr
&&+a^{\mu}\star\partial_{\mu}[(\mathcal{L}_{\star}\star e^{-1})e]
+\partial_{\mu}a^{\mu}\star(\mathcal{L}_{\star}\star e^{-1})e\Big\}.
\end{eqnarray}
Now by integrating on a submanifold $M \subset \mathbb{ R}^D$ with fields non vanishing at the boundary (so that the total derivative terms
 do not disappear),
 we get:
\begin{eqnarray}\label{noether}
\delta\mathcal{S}_{\star}= \int_{M}{\mbox{d}}^d x\hspace{1mm}
\partial_{\sigma}\Big[\mathcal{K}^{\sigma}+\mathcal{J}^{\sigma}
+a^{\sigma}\star\Big((\mathcal{L}_{\star}^{\Omega}\star e^{-1})e\Big)\Big]
\end{eqnarray}
coupled to the transformations
$
\delta\vp=-a^{\nu}\partial_{\nu}\vp,\quad\delta\vp^{c}=-a^{\nu}\partial_{\nu}\vp^{c},
\quad a^{\nu}=\mbox{constant},
$
that we substitute into (\ref{noether}) and taking into account the identities
$
\delta\vp^cX_c\partial_\mu\vp=\partial_\mu(\delta\vp^c X_c\vp)-\partial_\mu(\delta\vp^c e_c^\rho)\partial_\rho\vp
$
such that $\delta\vp^cX_c\partial_\mu\vp=\partial_\mu\delta\vp=-a^\nu\partial_\nu\partial_\mu\vp$ and the fact that 
 $e_{\nu}^{a}=\partial_{\nu}\vp^{a}$, we come from
the relation
\begin{eqnarray}
0=\delta\mathcal{S}_{\star}= -a^{\nu}\int_{M}{\mbox{d}}^d x\hspace{1mm}
 \partial_{\sigma}\mathcal{T}^{\sigma}_{\nu},
\end{eqnarray}
 where the EMT takes the form
\begin{eqnarray}\label{energy}
\mathcal{T}^{\sigma}_{\nu}&=&\frac{e}{2}\Big[(\partial_{\nu}\vp)\{\partial^{\sigma}\vp,e^{-1}\}_{\star}
+(\partial_{\nu}\vp_{c})\{\partial^{\sigma}\vp^{c},e^{-1}\}_{\star}\Big]\cr
&&+\frac{a}{2\theta^2}\Big[\partial_\sigma^{-1}(\partial_{\nu}\vp) \partial_\sigma^{-1}\Big(e\{\partial_\sigma^{-1}\vp,e^{-1}\}_\star\Big)+\partial_\sigma^{-1}(\partial_{\nu}\vp_c) \partial_\sigma^{-1}\Big(e\{\partial_\sigma^{-1}\vp^c,e^{-1}\}_\star\Big)\Big]
\cr
&&-ee_{b}^{\sigma}\Big[\mathcal{L}_{\star}\star(e^{-1}\partial_{\nu}\vp^{b})+T(\Delta)
\Big(X_{c}\mathcal{L}_{\star},\widetilde{X}^{b}(e^{-1}\partial_{\nu}\vp^{c})\Big)\Big].
\end{eqnarray}
This tensor is neither symmetric  and non locally conserved. Note that to recovered the EMT given in \eqref{emtpre}
we write 
$
T_{\nu\rho}=g_{\sigma\rho}\mathcal{T}^{\sigma}_{\nu}
$
and takes the limit $e_a^\mu\rightarrow \delta_a^\mu$. The expression \eqref{energy}  can be symmetrized as
\begin{eqnarray}\label{energy1211}
\mathcal{T}^{s}_{\nu\sigma}&=&\frac{e}{4}\Big[(\partial_{\nu}\vp)\{\partial_{\sigma}\vp,e^{-1}\}_{\star}
+(\partial_{\nu}\vp_{c})\{\partial_{\sigma}\vp^{c},e^{-1}\}_{\star}+(\partial_{\sigma}\vp)\{\partial_{\nu}\vp,e^{-1}\}_{\star}
+(\partial_{\sigma}\vp_{c})\{\partial_{\nu}\vp^{c},e^{-1}\}_{\star}\Big]\cr
&&+\frac{a}{4\theta^2}\Big[\partial_\sigma^{-1}(\partial_{\nu}\vp) \partial_\sigma^{-1}\Big(e\{\partial_\sigma^{-1}\vp,e^{-1}\}_\star\Big)+\partial_\sigma^{-1}(\partial_{\nu}\vp_c) \partial_\sigma^{-1}\Big(e\{\partial_\sigma^{-1}\vp^c,e^{-1}\}_\star\Big)\cr
&&+\partial_\nu^{-1}(\partial_{\sigma}\vp) \partial_\nu^{-1}\Big(e\{\partial_\nu^{-1}\vp,e^{-1}\}_\star\Big)+\partial_\nu^{-1}(\partial_{\sigma}\vp_c) \partial_\nu^{-1}\Big(e\{\partial_\nu^{-1}\vp^c,e^{-1}\}_\star\Big)\Big]
\cr
&&-\frac{e}{2}\Big\{ g_{\rho\sigma}e_{b}^{\rho}\Big[\mathcal{L}_{\star}\star(e^{-1}\partial_{\nu}\vp^{b})+T(\Delta)
\Big(X_{c}\mathcal{L}_{\star},\widetilde{X}^{b}(e^{-1}\partial_{\nu}\vp^{c})\Big)\Big]\cr
&&+g_{\rho\nu}e_{b}^{\rho}\Big[\mathcal{L}_{\star}\star(e^{-1}\partial_{\sigma}\vp^{b})
+T(\Delta)
\Big(X_{c}\mathcal{L}_{\star},\widetilde{X}^{b}(e^{-1}\partial_{\sigma}\vp^{c})\Big)\Big]\Big\}.
\end{eqnarray}

Now we can regularize the EMT \eqref{energy}. Due to the very complex form of expression in the general case we focus our attention 
 to the case where  the coordinates base $e_a^\mu(x)$ is to be $e_a^\mu=\delta_a^\mu+\omega_{ab}^\mu x^b$
such that the tensor $(\omega_{ab}^\mu)$ is symmetric 
between the index $a$ and $b$, i.e. $\omega_{ab}^\mu=\omega_{ba}^\mu$. The commutation relation between the vectors fields $X_a$ is:
\begin{eqnarray}
[X_a,X_b]=(\omega_{ba}^\mu-\omega_{ab}^\mu)\partial_{\mu}=0,
\end{eqnarray}
and therefore the dynamical star product is associative. We adequately choose the elements of the matrix $(\omega_{ab}^\mu)$ such that
 the matrix representation  of  $(e_a^\mu)$ is given in dimension $d=4$ by
\begin{eqnarray}
(e)_a^\mu=\left(\begin{array}{cccc} 
1+\omega_{11}^1x^1+\omega_{12}^1
x^2&\omega_{11}^2x^1+\omega_{12}^2x^2&0&0\\
\omega_{12}^1x^1+\omega_{22}^1x^2 &
1+\omega_{12}^2x^1+\omega_{22}^2 x^2&0&0\\
0&0&1+\omega_{33}^3x^3+\omega_{34}^3
x^4&\omega_{33}^4x^3+\omega_{34}^4x^4\\
0&0&\omega_{34}^3x^3+\omega_{44}^3x^4 &
1+\omega_{34}^4x^3+\omega_{44}^4 x^4
\end{array}\right).
\end{eqnarray}
Then, the determinants $e^{-1}$ and the inverse  $e$ becomes
\begin{eqnarray}
e^{-1}&=&1+\omega_\mu x^\mu,\quad e=1-\omega_\mu x^\mu,
\end{eqnarray}
where the components of the vector $\omega_\mu$ are
\bea
\omega_1=\omega_{11}^1+\omega_{12}^2,\quad\omega_2=\omega_{22}^2+\omega_{12}^1,\quad \omega_3=\omega_{33}^3+\omega_{34}^4,\quad \omega_4=\omega_{44}^4+\omega_{34}^3.
\eea
The noncommutative
tensor takes the form  $ \theta^{\mu\nu}(x)=\theta
e^{-1}J^{\mu\nu}$  where $(J)$ stands for the symplectic matrix in four dimensions.  
Besides,
the inverse matrix $e_{\mu}^a$ can be written as $
e_{\mu}^a=\delta_\mu^a+\omega^{ab}_\mu x_b,\,\,\, \mbox{ where
}\,\,\, \omega^{ab}_\mu=-\omega_{ab}^\mu ,$ and the solution of the
field equation $e_\mu^a=\partial_\mu\phi^a$ is well given by 
\bea
\phi^a=x^a+\frac{1}{2}\omega^{ab}_\mu x_b\, x^\mu.
\eea
Using all these considerations, after few algebraic computation we come to the relation
\bea
\partial^\nu\mathcal{T}^{s}_{\nu\sigma}&=&\frac{2e\lambda}{4!}X_aS(\Delta)([\partial_\sigma \vp,\vp]_\star,\widetilde{X}^a(\vp\star\vp\star e^{-1}))\cr
&=&\frac{2\lambda}{4!}\partial_\gamma\Big(e  e_a^\gamma S(\Delta)([\partial_\sigma \vp,\vp]_\star,\widetilde{X}^a(\vp\star\vp\star e^{-1}))\Big),
\eea
where the followings identities are used
\bea
&&\{\partial_\sigma\vp,e^{-1}\}_\star
=2e^{-1}\partial_\sigma\vp,\\ &&e^{-1}\partial_{\sigma}\vp^{c}=\delta_\sigma^c+\delta_\sigma^c \omega_\mu x^\mu+\omega_\sigma^{cd}x_d,\\
&&T(\Delta)
\Big(X_{c}\mathcal{L}_{\star},\widetilde{X}^{b}(e^{-1}\partial_{\sigma}\vp^{c})\Big)=0.
\eea
As the ordirary Moyal plane the EMT defined on the dynamical Moyal space can be regularized. We get the symmetric tensor
\bea
\mathcal{T}^{s,r}_{\nu\sigma}&=&\mathcal{T}^{s}_{\nu\sigma}-\frac{2\lambda}{4!}g_{\gamma\nu}\Big(e  e_a^\gamma S(\Delta)([\partial_\sigma \vp,\vp]_\star,\widetilde{X}^a(\vp\star\vp\star e^{-1}))\Big)\cr
&&-\frac{2\lambda}{4!}g_{\gamma\sigma}\Big(e  e_a^\gamma S(\Delta)([\partial_\nu \vp,\vp]_\star,\widetilde{X}^a(\vp\star\vp\star e^{-1}))\Big).
\eea
 By incorporating noncommutativity in the coordinates, the gravitation interaction can be taking into account in QFTs. However, the computation of the EMTs is based around
a prejudice for writing the Einstein field equations as $\mathcal G_{\nu\sigma}=\kappa \mathcal{T}^{s,r}_{\nu\sigma} $ with gravity on the left and matter on the
right. Due to the fact that $\partial^\nu\mathcal G_{\nu\sigma}=0$ we need improve the EMT such that $\partial^{\nu}\mathcal{T}^{s,r}_{\nu\sigma}=0$. 
As in \cite{Gerhold:2000ik}-\cite{AbouZeid:2001up} and \cite{Balasin:2015hna}we  have explicitly shown that the standard local conservation law of the EMT  is always modified due to non-commutative effects and that tensor can always be redefined so as to be conserved.

\section{Conclusion and remarks}\label{sec4}
 In conclusion, we summarize our results. We have
developed the variation techniques for the determination of the EL equations of motion of  a Lagrangian that depends on $\partial_\mu^{-1}\vp$. We have computed the EMT for the GMRT model, in ordinary and dynamical Moyal spaces.  The Wilson regularization procedure is also given to improve the corresponding tensors. 

Let us remark that

(i) The invariance of the action \eqref{action} under spacetime translation involves the locally conserved EMT, which needs not be symmetric and in a massless theory, it needs not be traceless either. The Lorentz symmetry in noncommutative theory is broken for $D>2$, since the constant skewsymmetric tensor $\theta^{\mu\nu}$ is not a Lorentz invariant tensor. As explaned in \cite{Aschieri:2008zv}, one way to restore this symmetry is to generalized the Moyal product defined by a set of a commuting vector field $X_a=e_a^\mu(x)\partial_\mu$. Therefore the EMT given with the translation invariant renormalizable action (The GMRT model) in the generalized Moyal space seems to be Lorentz invariant tensor. 

(ii) Introducing $x$-dependence in the deformation matrix $(\theta^{\mu\nu})$ of the star product leads to the definition of nontrivial background metric. Then the EMT associated to translation invariant field theory may provided from the core of the Einstein equation, when we assume that gravity can be  incorporated  in the noncommutativity.
Also, the EMT given in \eqref{energy12} can be regularized without  choosing  the tetrad as $e_a^\mu=\delta_a^\mu+\omega_{ab}^\mu x^b$.   In the general case of the tetrad $e_a^\mu(x)$, the same computation can be made easily, thanks to the example proposed in this paper.

\section*{Acknowledgements}
EB work is partially supported by
the  Abdus  Salam  International  Centre  for  Theoretical
Physics (ICTP, Trieste, Italy) through the Office of External Activities (OEA)-Prj-15.  The ICMPA is in partnership  with  the  Daniel  Lagolnitzer  Foundation  (DIF),
France.
DOS research at Max-Planck Institute  is supported by the Alexander von Humboldt foundation.

\begin{appendices}
\section{Formal variation principle of a nonlocal action $\mathcal L_\star$}
In this section we derive the variation principle in formal way, allows us to compute  the EMT \eqref{tensor1}. The nonlocal Lagrangian we consider here is of the form $\mathcal L_\star(\vp,\partial_\mu \vp,\partial_\mu^{-1}\vp)$. As mentioned in the section \eqref{sec2}, the nonlocality of $\mathcal L_\star$ comes from not only the star product, but also the inverse derivative of the field $\vp$ denoted by $\partial^{-1}\vp$. There are two ways for investigating this Lagrangian.  The first is to expand the star product and get a quantity of the form
\bea\label{nonloc}
\mathcal L_\star(\vp,\partial_\mu \vp,\partial_\mu^{-1}\vp)=\mathcal L(\vp, \partial_\mu\vp,\cdots,\infty,\partial^{-1}\vp,\theta).
\eea
The right hand side of this expression can be traited using the generalization of the Ostrogradski calculus for $n$ derivative Lagragian see \cite{Nakamura:1995qz}-\cite{Gomis:2000gy} for more explanation about this method. On the other hand, the left hand side may be   traited as  nonlocal noncommutative Lagrangian, and then the computation of the EMT is similar to what follows in \cite{Gerhold:2000ik}-\cite{Balasin:2015hna} apart the fact that the inverse derivative need to be considered carefully.  Let 
$S_\star[\vp]=\int\,d^dx\, \mathcal L_\star(\vp,\partial_\mu \vp,\Psi) $ is the nonlocal action coming from the nonlocal lagrangian \eqref{nonloc}, in which the quantity $\partial^{-1}\vp$ is considered as a new field denoted by $\Psi$: we get the  following variation:
\bea\label{mig2}
\delta S_\star[\vp]=\int\,d^dx\,\Big(\frac{\partial  \mathcal L_\star}{\partial\vp}\star\delta\vp+\frac{\partial  \mathcal L_\star}{\partial\partial_\mu\vp}\star\delta\partial_\mu\vp+\frac{\partial  \mathcal L_\star}{\partial\Psi}\star\delta\Psi\Big).
\eea
Now by considering  the following identities:
\bea
\partial_\mu \partial_\nu^{-1}\vp(x)=\delta_{\mu\nu}\vp(x)
,\quad \partial_\mu(a\star b)=(\partial_\mu a) \star b+a\star (\partial_\mu b),
\eea
we can simply deduce that
\bea
\frac{\partial  \mathcal L_\star}{\partial\partial_\mu\vp}\star\partial_\mu\delta\vp=\partial_\mu\Big(\frac{\partial  \mathcal L_\star}{\partial\partial_\mu\vp}\star\delta\vp\Big)-\partial_\mu\Big(\frac{\partial  \mathcal L_\star}{\partial\partial_\mu\vp}\Big)\star\delta\vp
\eea
\bea
 \frac{\partial \mathcal L_\star}{\partial\Psi}\star \delta\Psi=
\partial_\mu\Big[\partial_\mu^{-1}\Big(\frac{\partial \mathcal L_\star}{\partial\Psi}\Big)\star\delta\Psi\Big]-
 \partial_\mu^{-1}\Big(\frac{\partial \mathcal L_\star}{\partial\Psi}\Big)\star\delta\vp 1_\mu,
\eea
where the vector notation $1_\mu=\delta_{\mu\mu}$ is used to point out the fact that the Einstein summation holds. 
Then \eqref{mig2} becomes
\bea\label{labo}
\delta S_\star[\vp]&=&\int\,d^dx\,\Big(E_\vp\delta\vp+\partial^\mu J_\mu\Big),
\eea
where the quantities $E_\vp$ and $J_\mu$ are respectively given by
\bea
E_\vp:&=&\frac{\partial  \mathcal L_\star}{\partial\vp}-\partial_\mu\Big(\frac{\partial  \mathcal L_\star}{\partial\partial_\mu\vp}\Big)-
 \partial_\mu^{-1}\Big(\frac{\partial \mathcal L_\star}{\partial\Psi}\Big)1_\mu,\\ 
J_\mu:&=&\Big(\frac{\partial  \mathcal L_\star}{\partial\partial_\mu\vp}\star\delta\vp\Big)+\partial_\mu^{-1}\Big(\frac{\partial \mathcal L_\star}{\partial\Psi}\Big)\star \delta\Psi
\eea
The current $J_\mu$ involves the infinite number of derivatives respect to the field $\vp$ due to the definition of the star product,  also the inverse derivative of the form $\partial_\mu^{-1}\vp$, $\partial_\mu^{-2}\vp$ appear in this quantity.
Let $\partial^\alpha\vp\in\mathcal{S}(\mathbb{R}^d)$, $\alpha=\llbracket -2,-1 \rrbracket
\cup \mathbb{N}$, where $\mathcal{S}(\mathbb{R}^d)$ is the space of suitable Schwartzian functions. With this condition the surface term  vanish, i.e.  $ \int\,d^dx\,\partial^\mu J_\mu=0$ and the EL equation of motion $E_\vp=0$ is well satisfied. In the case of translation invariant action, the coordinates and field are transformed as: 
\bea
x'^{\mu}=x^\mu+a^\mu,\quad d^dx'=[1+\partial_\mu a^\mu+\mathcal O(a^2)]d^dx,\quad \delta\vp=-a^\mu\partial_\mu\vp,
\eea
such that the variation of the action $S_\star$ becomes ($a^\mu$ is a constant vector):
\bea
\delta  S_\star&=&\int\,d^dx \Big|\frac{\partial x'}{\partial x}\Big|\star \mathcal{L}'_\star-\int\, d^dx\,\mathcal{L}_\star\cr
&=&\int\,d^dx\Big(\delta_\vp \mathcal{L}_\star+a^\mu\partial_\mu \mathcal{L}_\star\Big)
\eea
Finally taking into account the fact that $E_\vp=0$  we come to
\bea
\delta  S_\star&=&\int\,d^dx\Big[\partial^\nu J_{\nu}(\delta\vp=-a_\mu\partial^\mu\vp)+a^\mu g_{\mu\nu}\partial^\nu\mathcal L_\star\Big]\cr
&=&-a^\mu\int\, d^dx\,\partial^\nu\Bigg[\Big(\frac{\partial  \mathcal L_\star}{\partial\partial_\nu\vp}\star\partial_\mu\vp\Big)+\partial_\nu^{-1}\Big(\frac{\partial \mathcal L_\star}{\partial\Psi}\Big)\star \partial_\mu\Psi- g_{\mu\nu}\mathcal L_\star\Bigg]\cr
&=&
-a^\mu\int\, d^dx\,\partial^\nu\Bigg[\frac{1}{2}\Big\{\frac{\partial  \mathcal L_\star}{\partial\partial_\nu\vp},\partial_\mu\vp\Big\}_\star+\frac{1}{2}\Big\{\partial_\nu^{-1}\Big(\frac{\partial \mathcal L_\star}{\partial\Psi}\Big), \partial_\mu\Psi\Big\}_\star- g_{\mu\nu}\mathcal L_\star\Bigg]\cr
&=&-a^\mu\int\, d^dx\, \partial^\nu T_{\nu\mu}.
\eea

\section{  Proof of relations \eqref{52} and \eqref{55}}
In this appendix we give the proof of the variation principle which leads to the EL equations of motion and the corresponding current and therefore contribute to the proof of relations \eqref{52} and \eqref{55}.
Consider the action \eqref{actiontanasa}.
Recall that the field $\vp_c$ do not depend for $\vp$. We get
\bea\label{48}
\delta_{\vp}S_{\partial}&=&\frac{a}{2\theta^2}\int\, e d^dx\,\Big[\partial_\mu^{-1}\delta\vp\star\partial_\mu^{-1}\vp\star e^{-1}+\partial_\mu^{-1}\vp\star\partial_\mu^{-1}\delta\vp\star e^{-1}\Big]\cr
&=&\frac{a}{2\theta^2}\int\, e d^dx\,\Big[\partial_\mu^{-1}\delta\vp\star\{\partial_\mu^{-1}\vp, e^{-1}\}_\star+2X_aS(\Delta)(\partial_\mu^{-1}\vp,\widetilde{X}^a(\partial_\mu^{-1}\delta\vp\star e^{-1}))\Big]\cr
&=&\frac{a}{2\theta^2}\int\, e d^dx\,\Big[(\partial_\mu^{-1}\delta\vp)\{\partial_\mu^{-1}\vp, e^{-1}\}_\star+ X_aT(\Delta)(\partial_\mu^{-1}\delta\vp,\widetilde{X}^a(\{\partial_\mu^{-1}\vp, e^{-1}\}_\star))\cr
&&+2X_aS(\Delta)(\partial_\mu^{-1}\vp,\widetilde{X}^a(\partial_\mu^{-1}\delta\vp\star e^{-1}))\Big],
\eea
where we have used the identities
\bea
\partial_\mu^{-1}\vp\star\partial_\mu^{-1}\delta\vp\star e^{-1}=\partial_\mu^{-1}\delta\vp\star e^{-1}\star\partial^{-1}_\mu\vp+2X_aS(\Delta)(\partial_\mu^{-1}\vp,\widetilde{X}^a(\partial_\mu^{-1}\delta\vp\star e^{-1}))\\
(\partial_\mu^{-1}\delta\vp)\star\{\partial_\mu^{-1}\vp, e^{-1}\}_\star=(\partial_\mu^{-1}\delta\vp)\{\partial_\mu^{-1}\vp, e^{-1}\}_\star+X_aT(\Delta)(\partial_\mu^{-1}\delta\vp,\widetilde{X}^a(\{\partial_\mu^{-1}\vp, e^{-1}\}_\star)).
\eea
Consider the following relation in which the repetitive indices in the right hand side are now summed:
\bea
 e(\partial_\mu^{-1}\delta\vp)\{\partial_\mu^{-1}\vp, e^{-1}\}_\star=\sum_{\mu}\partial_{\mu}\Big[(\partial_\mu^{-1}\delta\vp)\partial^{-1}_\mu (e\{\partial_\mu^{-1}\vp, e^{-1}\}_\star)\Big]-\delta\vp\, \partial^{-1}_\mu (e\{\partial_\mu^{-1}\vp, e^{-1}\}_\star).
\eea
By replacing this identity in \eqref{48},  we  get the $\vp$ variation of $S_{\partial}$ as 
\bea\label{54}
\delta_{\vp}S_{\partial}&=&\int\,d^dx\Big[\delta\vp E^\vp_\partial+\partial_\sigma\mathcal K^\sigma_\partial],
\eea
where $E_\partial$ contribute to the EL  equations of motion and $\mathcal K_\partial^\mu$ to the current:
\bea
E^\vp_\partial=-\frac{a}{2\theta^2}\partial^{-1}_\mu (e\{\partial_\mu^{-1}\vp, e^{-1}\}_\star),
\eea
\bea
\mathcal K^\sigma_\partial&=&\frac{a}{2\theta^2}\Big[(\partial_\sigma^{-1}\delta\vp)\partial^{-1}_\sigma(e\{\partial_\sigma^{-1}\vp, e^{-1}\}_\star)+  e e_b^\sigma T(\Delta)(\partial_\mu^{-1}\delta\vp,\widetilde{X}^b(\{\partial_\mu^{-1}\vp, e^{-1}\}_\star))\cr
&&+2  e e_b^\sigma S(\Delta)(\partial_\mu^{-1}\vp,\widetilde{X}^b(\partial_\mu^{-1}\delta\vp\star e^{-1}))\Big].
\eea
Using the same technical computation to  the remain expression of the functional action \eqref{action1} 
the EL equations of motion of the field $\vp$, i.e. $E^\vp=0$ and the corresponding current $\mathcal K^\sigma$ given repectively in the relations \eqref{EL50} and \eqref{C51} are well satisfied, and  then
\bea
\delta_{\vp}S_{\star}&=&\int\,d^dx\Big[\delta\vp E^\vp+\partial_\sigma\mathcal K^\sigma].
\eea

In the other hand we are interested to the $\vp_c$ variation of \eqref{actiontanasa}. This variation is sub-dived into two contributions namely $ A_\partial$ and $B_\partial$ such that
\bea
\delta_{\vp^c}S_\partial&=&\frac{a}{2\theta^2}\delta_{\vp^c}\left\{\int \,e d^dx\,\partial^{-1}_\mu \vp\star \partial^{-1}_\mu \vp\star e^{-1}\right\}+\frac{a}{2\theta^2}\delta_{\vp^c}\left\{\int \,e d^dx\,\partial^{-1}_\mu \vp_c\star \partial^{-1}_\mu \vp^c\star e^{-1}\right\}\cr
&=&A_\partial+B_\partial
\eea
where
\bea
A_\partial&=&\frac{a}{2\theta^2}\int\, d^dx\,(\delta_{\vp^c}e)\,(\partial^{-1}_\mu \vp\star \partial^{-1}_\mu \vp\star e^{-1})
+\frac{a}{2\theta^2}\int\, ed^dx\,\,\delta_{\vp^c}(\partial^{-1}_\mu \vp\star \partial^{-1}_\mu \vp\star e^{-1})\cr
&=&\frac{a}{2\theta^2}\int\, d^dx\,\left\{\partial_\sigma\Big(e e_a^\sigma\delta\vp^a\partial^{-1}_\mu\vp\star\partial_\mu^{-1}\vp\star e^{-1}\Big)- e e_a^\sigma\delta\vp^a\partial_\sigma(\partial_\mu^{-1}\vp\star\partial_\mu^{-1}\vp \star e^{-1})\right\}\cr
&+&\frac{a}{2\theta^2}\int\, ed^dx\,\Big\{-\delta\vp^a X_a\partial_\mu^{-1}\vp\{\partial_\mu^{-1}\vp,e^{-1}\}_\star-X_bT(\Delta)(\delta\vp^aX_a\partial_\mu^{-1}\vp,\widetilde{X}^b\{\partial_\mu^{-1}\vp,e^{-1}\}_\star)\cr
&-&2X_bS(\Delta)(\partial_\mu^{-1}\vp,\widetilde{X}^b\delta\vp ^aX_a\partial_\mu^{-1}\vp\star e^{-1})-X_a(\partial_\mu^{-1}\vp\star\partial_\mu^{-1}\vp \star \delta\vp^a e^{-1})\cr
&+&X_a(\partial_\mu^{-1}\vp\star\partial_\mu^{-1}\vp) \delta\vp^a e^{-1}+X_bT(\Delta)(X_a(\partial_\mu^{-1}\vp\star\partial_\mu^{-1}\vp),\widetilde{X}^b\delta\vp^a e^{-1})\cr
&+&\delta\vp^a X_a(\partial_\mu^{-1}\vp\star\partial_\mu^{-1}\vp\star e^{-1})\Big\}
\eea
and 
\bea
B_\partial&=&\frac{a}{2\theta^2}\int\,d^dx (\delta_{\vp^c}e)\Big(\partial^{-1}_\mu \vp_c\star \partial^{-1}_\mu \vp^c\star e^{-1}\Big)+\frac{a}{2\theta^2}\int\,ed^dx\, \delta_{\vp^c}\Big(\partial^{-1}_\mu \vp_c\star \partial^{-1}_\mu \vp^c\star e^{-1}\Big)\cr
&=&\frac{a}{2\theta^2}\int\,d^dx\,\Big\{\partial_\sigma\Big((e e_a^\sigma \delta\vp^a)\partial^{-1}_\mu \vp_c\star \partial^{-1}_\mu \vp^c\star e^{-1}\Big)-e e_a^\sigma\delta\vp^a\partial_\sigma(\partial^{-1}_\mu \vp_c\star \partial^{-1}_\mu \vp^c\star e^{-1})\Big\}\cr
&+&\frac{a}{2\theta^2}\int\, d^dx\Big\{-\delta\vp_c\partial_\mu^{-1}(e\{\partial_\mu^{-1}\vp^c,e^{-1}\}_\star)
+\partial_\mu\Big(\partial_\mu^{-1}\delta\vp_c\partial_\mu^{-1}(e\{\partial_\mu^{-1}\vp,e^{-1}\}_\star)\Big)\cr
&+&2\partial_\sigma\Big(e e_a^\sigma S(\Delta)(\partial_\mu^{-1}\vp_c,\widetilde{X}^a\delta\vp^c\star e^{-1})\Big)+\partial_\sigma\Big(e e_a^\sigma T(\Delta)(\partial_\mu^{-1}\delta\vp_c,\widetilde{X}^a\{\partial_\mu^{-1}\vp^c,e^{-1}\}_\star)\Big)\Big\}\cr
&+&\frac{a}{2\theta^2}\int\,ed^dx\,\Big\{-\delta\vp^aX_a\partial^{-1}_\mu\vp_c\{\partial^{-1}_\mu\vp^c,e^{-1}\}_\star-X_bT(\Delta)(\delta\vp^aX_a\partial^{-1}_\mu\vp_c,\widetilde{X}^b\{\partial^{-1}_\mu\vp^c,e^{-1}\}_\star)\cr
&-&2X_bS(\Delta)(\partial^{-1}_\mu\vp_c,\widetilde{X}^b\delta\vp^aX_a\partial^{-1}_\mu\vp^c\star e^{-1})-X_a(\partial^{-1}_\mu\vp_c\star \partial^{-1}_\mu\vp^c\star\delta\vp^a e^{-1})\cr
&+&\delta\vp^a e^{-1}X_a(\partial^{-1}_\mu\vp_c\star \partial^{-1}_\mu\vp^c)+X_bT(\Delta)(X_a(\partial^{-1}_\mu\vp_c\star \partial^{-1}_\mu\vp^c),\widetilde{X}^b\delta\vp^a e^{-1})\cr
&+&\delta\vp^a X_a(\partial^{-1}_\mu\vp_c\star \partial^{-1}_\mu\vp^c\star e^{-1})\Big\}.
\eea
Taking into account all of these quantities, the contribution to the EL equations of motion is
\bea
E^{\vp\vp^c}_{\partial}&=&\frac{a}{2\theta^2}\Big[-eX_a\partial_\mu^{-1}\vp\{\partial_\mu^{-1}\vp,e^{-1}\}_\star +X_a(\partial_\mu^{-1}\vp\star \partial_\mu^{-1}\vp)-eX_a\partial_\mu^{-1}\vp_c\{\partial_\mu^{-1}\vp^c,e^{-1}\}_\star \cr
&+&X_a(\partial_\mu^{-1}\vp_c\star \partial_\mu^{-1}\vp^c)
-\partial_\mu^{-1}(e\{\partial_\mu^{-1}\vp^c,e^{-1}\}_\star)\Big].
\eea
The contribution to the current $J^{\sigma}$ denote by $J^{\sigma}_\partial$ takes the form
\bea
J^{\sigma}_\partial&=&\frac{a}{2\theta^2}\Big[ee_b^\sigma\delta\vp^b\partial_\mu^{-1}\vp\star \partial_\mu^{-1}\vp\star e^{-1}-ee_b^\sigma T(\Delta)(\delta\vp^a X_a\partial_\mu^{-1}\vp,\widetilde{X}^b\{\partial_\mu^{-1}\vp,e^{-1}\}_\star))\cr
&&-2 ee_b^\sigma S(\Delta)(\partial_\mu^{-1}\vp,\widetilde{X}^b\delta\vp^a X_a\partial_\mu^{-1}\vp\star e^{-1})-e e_b^\sigma(\partial_\mu^{-1}\vp\star \partial_\mu^{-1}\vp\star \delta\vp^b e^{-1})\cr
&&+ee_b^\sigma T(\Delta)(X_a(\partial_\mu^{-1}\vp\star \partial_\mu^{-1}\vp),\widetilde{X}^b\delta\vp^a e^{-1})+e e_b^\sigma\delta\vp^b (\partial_\mu^{-1}\vp_c\star\partial_\mu^{-1}\vp^c\star e^{-1})\cr
&&+(\partial_\sigma^{-1}\delta\vp_c)\partial_\sigma^{-1}(e\{\partial_\sigma^{-1}\vp^c,e^{-1}\}_\star)+2 e e_b^\sigma S(\Delta)(\partial_\mu^{-1}\vp_c,\widetilde{X}^b\delta\vp^c e^{-1})\cr
&&+ee_b^\sigma T(\Delta)(\partial_\mu^{-1}\delta\vp_c,\widetilde{X}^b\{\partial_\mu^{-1}\vp^c,e^{-1}\}_\star)-ee_b^\sigma T(\Delta)(\delta\vp^a X_a\partial_\mu^{-1}\vp_c,\widetilde{X}^b(\{\partial_\mu^{-1}\vp^c,e^{-1}\}_\star))\cr
&&-2e e_b^\sigma S(\Delta)(\partial_\mu^{-1}\vp_c,\widetilde{X}^b\delta\vp^a X_a\partial_\mu^{-1}\vp^c\star e^{-1})-ee_b^\sigma(\partial_\mu^{-1}\vp_c\star \partial_\mu^{-1}\vp^c\star\delta\vp^a e^{-1})\cr
&&+ee_b^\sigma T(\Delta)(X_a(\partial_\mu^{-1}\vp_c\star \partial_\mu^{-1}\vp^c),\widetilde{X}^b\delta\vp^a e^{-1})\Big].
\eea
By performing the same computation to the other terms in the action \eqref{action1} we get the EL equations of motion \eqref{EL53}
and the current \eqref{C54}, 
such that 
\begin{eqnarray}
\delta_{\vp^{c}}\mathcal{S}_{\star}&=&\int{\mbox{d}}^d x\hspace{1mm}\Big(\delta\vp^{c}\,{E}^{\vp\vp^c}+\partial_{\sigma}\mathcal{J}^{\sigma}\Big).
\end{eqnarray}
\end{appendices}

\end{document}